\documentclass[12pt]{article}

\usepackage{placeins}
\usepackage{pifont}
\usepackage{graphicx}
\usepackage{booktabs}
\usepackage{multirow}
\usepackage{amsmath}
\usepackage{amssymb}
\usepackage{lmodern}
\usepackage{rotating}
\usepackage{float}
\usepackage{graphicx}
\usepackage{xcolor}
\usepackage{listings}
\usepackage{framed}

\usepackage{tabularx}
\usepackage{array}
\usepackage{makecell}

\usepackage[T1]{fontenc}
\usepackage[a4paper,left=2.5cm,right=2.5cm,top=3cm,bottom=3cm]{geometry}
\usepackage[flushleft]{threeparttable}

\usepackage[style=authoryear-comp, backend=bibtex, maxcitenames=2, maxbibnames=99 ,dashed=false]{biblatex}
\bibliography{refs_adjustedCurves}

\usepackage{hyperref}

\newcolumntype{L}{>{\raggedright\arraybackslash}p{0.36\textwidth}}
\newcolumntype{C}{>{\centering\arraybackslash}p{0.16\textwidth}}

\newcommand{\bigCI}{\mathrel{\text{\scalebox{1.07}{$\perp\mkern-10mu\perp$}}}}

\definecolor{codegreen}{rgb}{0,0.6,0}
\definecolor{codegray}{rgb}{0.5,0.5,0.5}
\definecolor{codepurple}{rgb}{0.58,0,0.82}
\definecolor{backcolour}{rgb}{0.95,0.95,0.92}

\definecolor{rstatement}{HTML}{cd6932}
\definecolor{rlogical}{HTML}{af3e53}

\lstdefinestyle{mystyle}{
	language=R,
	backgroundcolor=\color{backcolour},    
	commentstyle=\color{codegreen},
	keywordstyle=\color{rstatement},
	numberstyle=\tiny\color{codegray},
	stringstyle=\color{codegreen},
	basicstyle=\ttfamily\normalsize,
	tabsize=2,
	morekeywords={adjustedsurv, adjusted_curve_diff, plot_curve_diff, adjusted_surv_quantile, adjusted_rmst, coxph, Surv},
	keywordstyle=[3]{\color{rlogical}},
	morekeywords=[3]{TRUE, FALSE},
	otherkeywords={!,!=,~,$,*,\&,\%/\%,\%*\%,\%\%,<-,<<-,/},
	alsoother={.$},
	deletekeywords={data, variable, na, action, legend, title, family},
	numbers=left,                    
	numbersep=5pt,
}

\begin{document}

\title{adjustedCurves: Estimating Confounder-Adjusted Survival Curves in R}
\date{}
\author{Robin Denz and Nina Timmesfeld \\ \\ Ruhr-University Bochum \\ Department of Medical Informatics, Biometry and Epidemiology}

\maketitle

\begin{abstract}
	Kaplan-Meier curves stratified by treatment allocation are the most popular way to depict causal effects in studies with right-censored time-to-event endpoints. If the treatment is randomly assigned and the sample size of the study is adequate, this method produces unbiased estimates of the population-averaged counterfactual survival curves. However, in observational studies, this is no longer the case. Instead, specific methods that allow adjustment for confounding must be used. We present the \texttt{adjustedCurves} \textbf{R} package, which can be used to estimate and plot these confounder-adjusted survival curves using a variety of methods from the literature. It provides a convenient wrapper around existing \textbf{R} packages on the topic and adds additional methods and functionality on top of it, uniting the sometimes vastly different methods under one consistent framework. Among the additional features are the estimation of confidence intervals, confounder-adjusted restricted mean survival times and confounder-adjusted survival time quantiles. After giving a brief overview of the implemented methods, we illustrate the package using publicly available data from an observational study including 2982 breast cancer.
\end{abstract}

\section{Introduction} \label{sec:intro}

Treatment-specific survival curves are the standard way to visualize causal treatment effects in studies with time-to-event endpoints. The Kaplan-Meier estimator \parencite{Kaplan1958} is commonly used to estimate these curves. In large randomized controlled trials with a balanced distribution of confounders across groups, this estimator produces unbiased estimates. However, this is not the case in observational studies, where the treatment was not assigned randomly or in randomized studies, when some of the participating individuals did not adhere to their randomly assigned treatment. In such studies, confounders of the treatment-outcome relationship can systematically bias the depicted survival curves \parencite{Nieto1996}.
\par
Although confounding is a well-known problem, it is often ignored when estimating survival curves. In many cases, the main analysis is carried out using a suitable regression model, such as the Cox proportional hazards model \parencite{Cox1972}. The estimated adjusted hazard ratios are then reported in tabular form. For graphical depiction, only simple un-adjusted Kaplan-Meier curves are used \parencite{Dey2020}. This is problematic, because the adjusted regression analysis may produce different results than the un-adjusted analysis if confounding is present. In this case, the depiction of the treatment effect in the un-adjusted graphs will likely not match the regression analyses, confusing the reader \parencite{Hu2020}.
\par
One possible way to avoid this confusion is to not display any visualization of the treatment effect at all. This approach is flawed, because graphics facilitate the communication of statistical analyses \parencite{Davis2010, Zipkin2014, Lundgreen2021}. This is especially important when the aim of the study is to estimate causal effects. As \textcite[p. 482]{Cunningham2021} aptly put it: "causal inference studies desperately need data visualization of the main effects".
\par
Another reason not to abstain from survival curve estimation is that they can easily be used to obtain absolute measures of effect, while time-to-event-based regression models by themselves usually only result in relative effect measures, such as the hazard ratio. For example, adjusted survival curves can be used to directly estimate marginal treatment effects using differences at specific times \parencite{Klein2007}, restricted mean survival times \parencite{Conner2019} or survival time quantiles \parencite{Chen2016}. Both the CONSORT Guidelines (``Consolidated Statement of Reporting Trials'') \parencite{Schulz2010} and the STROBE Statement (``Strengthening the Reporting of Observational Studies in Epidemiology'') \parencite{Vandenbroucke2007} recommend the use of such absolute effect measures.
\par
Luckily, multiple methods to estimate confounder-adjusted survival curves have been proposed in the literature. These can roughly be divided into \emph{Stratification} based methods \parencite{Nieto1996, Cupples1995, Gregory1988, Amato1988}, \emph{G-Computation} \parencite{Makuch1982, Chang1982}, \emph{Inverse Probability of Treatment Weighting} (IPTW) \parencite{Cole2004, Xie2005}, \emph{Propensity Score Matching} \parencite{Austin2014}, \emph{Empirical Likelihood Estimation} (EL) \parencite{Wang2019}, \emph{Augmented Inverse Probability of Treatment Weighting} (AIPTW) \parencite{Robins1992, Ozenne2020}, \emph{Targeted Maximum Likelihood Estimation} (TMLE) \parencite{Moore2009, Stitelman2010, Cai2020}, an instrumental-variable based method \parencite{MartinezCamblor2020}, proximal causal inference based methods \parencite{Ying2022} and some of their \emph{Pseudo-Values} (PV) based versions \parencite{Andersen2017, Wang2018}. A small review and extensive simulation study is given by \textcite{Denz2022}.
\par
There is currently no single software package which implements all of these methods in a consistent framework. Existing implementations focus only on one or a few methods or offer no suitable plotting routines. The \texttt{adjustedCurves} \textbf{R} package directly addresses these deficits by providing high-level user-friendly implementations of all methods named above (with the exception of TMLE based methods), including confidence interval calculation. It also provides highly-customizable \texttt{plot()} functions based on the \texttt{ggplot2} package \parencite{Wickham2016} to produce publication-ready graphics. The package is designed to be usable by researchers with limited \textbf{R} experience, while still offering a substantial amount of arguments and functionality for more advanced users. A stable release version is available on CRAN \parencite{Denz2022a}, while the developmental version can be found on github (\url{https://github.com/RobinDenz1/adjustedCurves}). This article presents this package.
\par
First, we formally define what confounder-adjusted survival curves are and describe ways to estimate marginal treatment effects based on them. Afterwards, we give a short review of all implemented methods to estimate confounder-adjusted survival curves from observational data. Next, we review some of the existing software implementations in the most popular programming languages \textbf{R}, \textbf{Stata} and \textbf{SAS}. Subsequently, we describe the concrete implementation in the proposed \texttt{adjustedCurves} package and give some simple examples of its usage based on the \texttt{rottderam} data set \parencite{Foekens2000} contained in the \texttt{survival} \textbf{R} package \parencite{Therneau2021}. Finally, we give a brief discussion about the proposed \textbf{R} package.

\section{Target Estimand} \label{sec::target_estimand}

Let $T$ be the time until the occurrence of the event of interest. Let $Z \in \{0, 1, ..., k\}$ be a treatment indicator, where each integer represents one of $k$ possible treatments at baseline. Under the potential outcomes framework, there are $k$ potential survival times $T^{z}$, one for each of the $k$ possible treatments \parencite{Neyman1923, Rubin1974}. We further define $X$ to be a vector of baseline covariates for each individual $i$, where $(T_i, Z_i, X_i), i = 1, ..., n$ are independent observations. The conditional counterfactual survival probability of is then given by:

\begin{equation}
	S_z(t | x) = P(T^{z} > t | X = x),
\end{equation}

where $T_{z}$ is the survival time that would have been observed if $Z$ was set to $z$ by some intervention \parencite{Cai2020, Woodward2003}. Although this conditional quantity may be very useful in some cases, it is not the primary target estimand here. Instead, all methods discussed here aim to estimate the marginal survival probability that would have been observed in some defined super-population, if every individual in that population had received treatment $z$. It can be formally defined as:

\begin{equation} \label{eq::target_estimand}
	S_{z}(t) = P(T^z > t) = E(S_z(t, X)).
\end{equation}

This quantity is known as the the population level \emph{counterfactual survival probability}, sometimes also called \emph{causal survival probability} or \emph{confounder-adjusted survival probability}. We use all three terms interchangeably in this paper. When this function is displayed over time, it is often called the \emph{counterfactual survival curve} etc. \parencite{Denz2022}. As a population-level quantity, it is often more useful than the conditional survival probability for healthcare planners and decision makers, who aim to maximize survival in a target population. It also more closely resembles what is usually estimated in high-quality randomized controlled trials \parencite{Hernan2016}.
\par
In order to estimate $S_z(t)$ in practice, four causal identifiability assumptions have to be met. The first one is the \emph{conditional exchangeability} (also known as \emph{no unmeasured confounding}) assumption, which states that, inside the strata defined by the relevant confounders, the outcome of the study would not change if the treatment groups were exchanged. Formally, this can be written as $T^{z} \bigCI Z | X$ marginally for each $z \in \{0, 1, ..., k\}$ \parencite{Sarvet2020}. The second assumption is known as the \emph{positivity} assumption, which states that the probability of receiving each possible treatment must be higher than 0 and lower than 1 for every individual \parencite{Hernan2019}. Given the observed confounders $X$, this can be represented as $P(Z = z | X) > 0$ and $P(Z = z | X) < 1$ for all possible values of $Z$.
\par
The third assumption is that the potential survival time would have been the same regardless of how the treatment status was set to $Z$. This is known as the \emph{counterfactual consistency} assumption and may be formally stated as $T_i = T_i^z$ if $Z_i = z$ for all $i$ and $z$ \parencite{Hernan2016}. Consistency may be violated if the treatment is not 'well-defined', meaning that there are hidden variations of the treatment $z$ that would lead to a different potential survival time \parencite{VanderWeele2018}. Finally, the \emph{no interference assumption} has to hold. This means that the potential survival time of each person depends only on its own treatment assignment and is unaffected by the treatment assignment of other individuals \parencite{Naimi2015}. An in-depth discussion of these assumptions can be found in \textcite{Hernan2019} and \textcite{Guo2015}.
\par
Additionally, in studies with time-to-event endpoints, \emph{right-censoring} is a frequent issue. Right-censoring means that, instead of observing $T$ directly, we only observe $T^{obs}$ with a corresponding event indicator $D$, where $T^{obs} = \min (T, C)$ and $D = I(T \geq C)$, with $C$ being the observed censoring time. In order to identify~\ref{eq::target_estimand} from such censored data, we need to make two further identifiability assumptions. The first can be stated as $T \bigCI C | Z, X$, which states that within each treatment group, $X$ renders censoring independent of the survival time. The second assumption states that the conditional probability of remaining uncensored up to $t$, defined as $P(C > t | Z = z, X = x)$ must be positive at each level of the covariate $x$.

\section{Defining Average Treatment Effects}

Visualizing estimates of $S_z(t)$ for all possible $Z$ over the observed range of time is a great way to visualize the treatment effect of $Z$ on the survival time. However, it may also be useful to compare these counterfactual survival curves more directly. This can be done by directly contrasting them or by first deriving statistics that summarize $S_z(t)$ in some way and contrasting those. In this way, \emph{marginal treatment effects} (also known as \emph{marginal causal effects}) are defined directly on the basis of the target estimand.\footnote{Depending on the considered target population for this estimand, different kinds of effects can be targeted. When the target population includes all individuals, irrespective of their observed treatment status, the results are denoted as average treatment effects (ATE). When only considering the treated or untreated sub-populations, the results are denoted as average treatment effect on the treated (ATT) or average treatment effect on the untreated (ATC) instead \parencite{Hernan2019}.} The \texttt{adjustedCurves} package offers three such statistics: the difference between counterfactual survival probabilities, the counterfactual survival time quantiles and the counterfactual restricted mean survival time. All of these can be estimated by utilizing estimates of $S_z(t)$ directly.
\par
The advantage of using these summary statistics instead of more popular effect measures, such as the hazard ratio obtained from a Cox model, is that they are absolute measures of effect that can be interpreted causally (under the assumptions stated above). In contrast, it has been shown repeatedly that the hazard ratio obtained by using a Cox model does not allow causal interpretations, not even in randomized controlled trials \parencite{Hernan2010, Aalen2015}. Using the effect measures presented here instead, may thus enhance the causal interpretability of the results. Since the effect measures are directly derived from survival probability estimates, any adjustment method may be used to estimate them, which additionally allows researchers more flexibility.

\subsection{Differences in Survival Probabilities}

The simple difference between two treatment-specific counterfactual survival probabilities at some specified point in time $t$ is given by:

\begin{equation}
	\Delta_{a,b}(t) = S_{z=a}(t) - S_{z=b}(t),
\end{equation}

where $a$ and $b$ are some possible values of $Z$. This difference is an absolute measure of effect and has a very straightforward interpretation. It is, however, a time-dependent quantity and as such requires a theoretically grounded choice of $t$ if one wants to use it to define the average treatment effects \parencite{Klein2007, Wang2019}. Alternatively, one can depict estimates of $\Delta_{a,b}(t)$ over time to highlight differences between two treatment groups \parencite{Coory2014}. If a relative effect measure is desired, the ratio of the survival probabilities may of course also be calculated.

\subsection{Survival Time Quantiles}

Another well known statistic to compare two survival functions is the \emph{median survival time} \parencite{Chen2016, BenAharon2019}. It is defined as the first point in time at which the survival probability in the respective group is equal to or below 0.5. Instead of using the median, other quantiles such as the 0.25 quantile or the 0.75 quantile may also be used. Formally, such survival time quantiles are defined as:

\begin{equation}
	\hat{Q}_z(p) = min\left(t | S_z(t) \leq p\right),
\end{equation}

where $p$ is the survival time quantile of interest. In other words, this quantity can simply be read off $S_z(t)$ by drawing a horizontal line at $S_z(t) = p$ and taking the first associated value of the survival time $t$ at this point. The difference or ratio between the quantiles for different values of $z$ may also be used to define marginal treatment effects.

\subsection{Restricted Mean Survival Times}

The counterfactual restricted mean survival time is defined as the area under $S_z(t)$ from 0 to some previously specified point in time $\tau$ \parencite{Conner2019}. Formally, the RMST is defined as:

\begin{equation}
	\text{RMST}_z(\tau) = \int_{0}^{\tau} S_{z}(t) dt.
\end{equation}

It can be interpreted as the mean survival time of the individuals in that interval. Multiple authors advocate for its use, naming the natural interpretation and ease of estimation as desirable properties \parencite{Royston2013, Hasegawa2020}. As with the survival time quantiles, one may use the difference or ratio between two treatment-specific $\text{RMST}_z(\tau)$ values to define the average causal effect \parencite{Ambrogi2020}. Although this quantity is a summary measure of $S_z(t)$ over a specific range of time, it is still dependent on the choice of $\tau$. Some guidance to the choice of $\tau$ has been published \parencite{Sheldon2013} but it remains a difficult endeavor.
\par
Instead of choosing a single $\tau$, it is also possible to estimate $\text{RMST}_z(\tau)$ for a range of $\tau$ values and plot the results over time \parencite{Zhao2016}. All of these options are also implemented in the \texttt{adjustedCurves} package.

\subsection{Choosing a Summary Measure}

The three summary measures described above answer different questions, and no single measure is universally preferable. The difference in survival probabilities at a prespecified time point is particularly useful when the clinical or scientific question concerns the probability of being event-free at a specific, meaningful landmark \parencite{Klein2007}. It is therefore often easy to communicate to practitioners and patients, but it only reflects the survival probabilities at the selected time point and does not summarize what happened earlier during follow-up. Survival time quantiles, such as the median, are useful when the time until a specified proportion of individuals experience the event is of primary interest and provide an intuitive interpretation on the time scale. However, they require sufficient follow-up to estimate the corresponding quantile and may be relatively insensitive to differences occurring away from that particular point of the survival distribution \parencite{Chen2016}.
\par
In contrast, RMST summarizes the entire survival experience up to a prespecified time horizon and its difference can be interpreted as the average amount of event-free time gained or lost during that period. This makes RMST particularly attractive when survival curves are non-proportional, cross, or differ over extended periods of follow-up. Its main limitation is that the interpretation depends on the chosen time horizon $\tau$, which should ideally be selected based on clinical or scientific considerations rather than the observed data.
\par
Consequently, the choice of summary measure should be guided by the scientific question: fixed-time survival differences are most natural for landmark questions, quantiles for questions about the timing of events in a specified part of the survival distribution, and RMST for questions concerning the average event-free time over a defined follow-up period \parencite{Royston2013, Hasegawa2020, Averbuch2025}.

\section{Methods} \label{sec::methods}

In the following section we will give a short review of all methods that may be used to estimate $S_z(t)$ using the \texttt{adjustedCurves} package, if the stated identifiability assumptions are met. The \texttt{adjustedCurves} package focuses strictly on binary or categorical treatments that do not change over time. Methods to estimate $S_z(t)$ for continuous $Z$ \parencite{Denz2023} or for time-varying $Z$ \parencite{Clare2019} are therefore not mentioned here. For simplicity of presentation, we will assume that right-censoring is completely at random, meaning that the probability of right-censoring is independent of $X$. Some methods in this package directly support right-censoring dependent on $X$, as discussed were appropriate. Other forms of censoring, such as left-censoring or interval-censoring \parencite{Kleinbaum2012} are not supported by the \texttt{adjustedCurves} package and are therefore not discussed here. However, all methods presented here also work if the data contains no right-censoring.
\par
As stated above, we generally assume conditional exchangeability, if not stated otherwise. This means that we require all confounders to be known and measured, which is very difficult in practice. To identify the relevant confounders, causal directed acyclic graphs may be used in conjunction with Pearls backdoor criterion \parencite{Pearl2009}. If some of the identified confounders are not measured, most methods included in this package are not suitable to obtain unbiased estimates of $S_z(t)$. The only exception are the proximal causal inference methods introduced by \textcite{Ying2022}, which, however, require a specific causal structure of the underlying data generation process, as discussed below. Furthermore, the instrumental-variable based strategie described by \textcite{MartinezCamblor2020} may be useful under some forms of unmeasured confounding. It does, however, not estimate $S_z(t)$ directly. Instead, it estimates the same population-averaged survival probability among a specific subgroup of compliers, as also discussed below. Both the proximal causal inference and instrumental variable based methods are also included in the \texttt{adjustedCurves} package. For all other methods we will assume that the relevant confounders have been identified and are fully measured without error.
\par
For the sake of simplicity, we will not give detailed equations for all estimators implemented in this \textbf{R} package. Instead we only present some equations where necessary or useful and give a short text description of how each method works otherwise. The full equations of the estimators and, if applicable, their respective variance estimators, can be found in the cited literature.

\subsection{No Adjustment} \label{sec::methods_crude}

The standard Kaplan-Meier (KM) estimator \parencite{Kaplan1958} of the un-adjusted survival function for treatment $z$ is given by:

\begin{equation}
	\hat{S}_z(t) = \prod_{j: t_j \leq t} \left(1 - \frac{d_{jz}}{n_{jz}}\right),
\end{equation}

where $d_{jz}$ is the number of events at time $t_j$ in treatment group $z$ and $n_{jz}$ is the number of individuals still at risk up to time $t_j$ in group $z$, that is all individuals who did not experience an event previously and who are not censored. As stated before, this $\hat{S}_z(t)$ is not an unbiased estimator of $S_{z}(t)$ if confounding is present. Within our package, this estimator can be calculated using \texttt{method="km"} in the \texttt{adjustedsurv()} function. More details on the implementation is given below. These estimators are included for reference only and should not be used when confounder-adjustment is the goal of the analysis.

\subsection{Stratification} \label{sec::methods_strat}

A very simple way to remove confounding is by performing an analysis stratified by each possible strata of the confounders and calculating an appropriately weighted pooled estimate afterwards. If there are only a few confounders which happen to have a small number of possible values, this is a valid strategy. For survival curve estimation, we can calculate simple Kaplan-Meier estimates stratified by the confounders and treatment, and take a weighted average of these Kaplan-Meier estimates inside each level of $Z$ at some point in time $t$. The weights are simply the number of occurrences of each confounder level in the full sample \parencite{Kramar1990, Cupples1995}.
\par
This method is implemented in the \texttt{adjustedsurv()} function under \\ \texttt{method="strat\_cupples"}. A very similar estimator that was proposed independently by \textcite{Gregory1988} and \textcite{Nieto1996} is also implemented in this function using \texttt{method="strat\_nieto"}. Both methods, however, only work for those $t$ in which there is still at least one person at risk in each stratum. To overcome this difficulty, \textcite{Amato1988} proposed a more sophisticated version of this estimator (\texttt{method="strat\_amato"}).
\par
Although these methods can only be used with a limited number of categorical confounders, they should not be disregarded. In cases with only a few categorical confounders, they offer a simple and, most importantly, non-parametric way to estimate the quantity of interest. Most other methods implemented in this package rely on the correct specification of at least one parametric or semi-parametric model, which may be difficult in practice.

\subsection{G-Computation} \label{sec::methods_direct}

This method is also known as \emph{direct standardization}, the \emph{corrected-group prognosis} method, or more generally, as a \emph{g-formula} method \parencite{Makuch1982, Chang1982, Robins1986}. First, a model describing the time-to-event outcome is estimated using the available data. Let this model be defined as:

\begin{equation}
	m_z(t, x) = P(T > t | Z = z, X = x).
\end{equation}

This model should include all relevant confounders as independent covariates $X$ and should directly account for right-censoring. The Cox-Proportional hazards model is often used, although many other suitable model may be used as well. The fitted model is then used to make predictions for the conditional probabilities for every person $i$ under treatment $z$ at $t$. The estimator for $S_{z}(t)$ is then defined as:

\begin{equation}
	\hat{S}_z(t) = \frac{1}{n} \sum_{i = 1}^{n} \hat{m}_z(t, X_i),
\end{equation}

where $n$ is the total sample size and $\hat{m}_z(t, X_i)$ are the conditional survival predictions. To make these kind of predictions using a Cox model, an additional estimate of the baseline-hazard function is needed, which can be estimated directly from the data \parencite{Breslow1972}. In general, any model that produces conditional survival probability predictions may be used. Possibly alternatives are additive hazards models \parencite{Aalen1980} or flexible hazards models \parencite{Royston2002}. This is directly supported in the \texttt{adjustedCurves} package.
\par
In addition to the four identifiability assumptions, this method also requires that the time-to-event model is correctly specified, which means that all assumptions made when fitting the model have to be met as well. If this is not the case, the resulting estimates are biased. In our package, g-computation can be performed using \texttt{method="direct"} in the \texttt{adjustedsurv()} function. It internally relies on the fast \texttt{predictRisk()} function of the \texttt{riskRegression} \parencite{Gerds2021} package to obtain the conditional survival probability predictions.

\subsection{Inverse Probability of Treatment Weighting (IPTW)} \label{sec::methods_iptw}

Instead of modelling the time-to-event outcome, we can also adjust for confounding using the \emph{treatment-assignment} mechanism. This usually entails fitting a model to predict the conditional probability of receiving treatment $Z$, defined as:

\begin{equation}
	\pi(Z) = P(Z | X = x),
\end{equation}

which is formally known as the \emph{propensity score}. By conducting an analysis weighted by the inverse of this probability $w(Z) = 1/\pi(Z)$, the distribution of the confounders is approximately equal between the treatment groups and the confounding is removed. That is, if $\pi(Z)$ was correctly estimated. Popular models to estimate this quantity are logistic regression models when $Z$ is binary and multinomial regression models when $Z$ is categorical. It is, however, not necessary to rely on the propensity score. Other weights $w(Z)$ that can be used to create covariate balance between treatment groups have been proposed in the literature \parencite{Imai2014, Hainmueller2012}. Those are also supported by the \texttt{adjustedCurves} package. 
\par
After such balancing weights have been estimated, the following estimator can be used to estimate $S_{z}(t)$ as:

\begin{equation}
	\hat{S}_z(t) = \left\{
	\begin{array}{ll}
		1  & \mbox{if } t < t_1 \\
		\prod_{t \leq t_j} \lbrack 1 - d_{jz}^w / Y_{jz}^w \rbrack & \mbox{if } t \geq t_1
	\end{array}
	\right.
\end{equation}

for $Y_{jz}^w > 0$, where $Y_{jz}^w$ is the weighted number of individuals at risk in group $z$, defined as:

\begin{equation}
	Y_{jz}^w = \sum_{i: T_i^{obs} \geq t_j} w(Z_i) I(Z_i = z),
\end{equation}

and $d_{jz}^w$ is defined as the weighted number of individuals with events out of the $Y_{jz}^w$ people at risk in group $z$, given by:

\begin{equation}
	d_{jz}^w = \sum_{i: T_i^{obs} = t_j} w(Z_i) A_i I(Z_i = z).
\end{equation}

This estimator is implemented in our package as \texttt{method="iptw\_km"} in the \\ \texttt{adjustedsurv()} function. A very similar estimator that has been proposed by \textcite{Cole2004} is also implemented in that function (\texttt{method="iptw\_cox"}). Note that, similar to the classic Kaplan-Meier estimator, this estimator does not directly support dependent censoring. If censoring depends on $X$, other methods, such as g-computation, may be more suitable. If users want to stick with an inverse probability weighting based estimator, the Pseudo-value based approach discussed in section~\ref{sec::methods_pseudo}, with inverse probability of censoring weighted pseudo-values, may also be used.

\subsection{Propensity Score Matching} \label{sec::methods_matching}

A different way to utilize the propensity score for confounder-adjustment is \emph{Propensity Score Matching} \parencite{Austin2014}. By matching individuals with similar propensity scores to each other, a sample is generated in which the propensity scores are equally distributed between treatment groups. This procedure creates covariate balance, if the model used to estimate the propensity scores is correctly specified and the matching algorithm is appropriate. We can then use standard methods, such as the Kaplan-Meier estimator, on the matched data to get an unbiased estimates of the causal survival curve. This estimator can be used by setting \texttt{method="matching"} in the \texttt{adjustedsurv()} function. Internally, it relies on the \texttt{Matching} package \parencite{Sekhon2011} to construct the matched data set using nearest neighboor matching with replacement.
\par
There are other methods to obtain matching based estimates of the causal survival curve described in the literature \parencite{Winnett2002, Galimberti2002, Austin2020}. We choose not to implement those methods directly, because of the large amount of modeling and parameter choices that need to be made to obtain valid estimates. We believe that there is no way to automate these methods without enticing bad research practice. The matching estimator here is included primarily as a simple example and to raise awareness that matching is a valid alternative to the other implemented methods.

\subsection{Empirical Likelihood Estimation (EL)} \label{sec::methods_emp_lik}

The previous two methods relied on an unbiased estimate of the propensity score, which, as mentioned above, are usually estimated using parametric models such as the logistic regression model. When using parametric models, assumptions need to be made about the functional form of the relationship between confounders and the treatment variable. To circumvent the need for these assumptions, \textcite{Wang2019} proposed the use of \emph{Empirical Likelihood Estimation} (EL) to obtain an estimate of the causal survival curve. This method is very similar to the IPTW estimator described in section~\ref{sec::methods_iptw}, but the weights are obtained using EL. In EL, a constrained likelihood function is maximized which forces the moment of the confounders to be approximately equal between treatment groups.
\par
This estimator is available for causal survival curves using \texttt{method="emp\_lik"} in the \texttt{adjustedsurv} function. It relies on code from the \texttt{adjKMtest} package (available at \url{https://github.com/kimihua1995/adjKMtest}). An alternative way to get similar estimates is to estimate non-parametric covariate balance weights \parencite{Chan2016} using the \texttt{WeightIt} package \parencite{Greifer2021} and passing those to an IPTW estimator such as the one described in section~\ref{sec::methods_iptw} (\texttt{method \%in\% c("iptw\_km", "iptw\_cox", "iptw\_pseudo")}).

\subsection{Augmented Inverse Probability of Treatment Weighting \\ (AIPTW)} \label{sec::methods_aiptw}

All methods mentioned above relied either on creating covariate balance between the treatment groups (IPTW, Propensity Score Matching, EL) or on directly substituting the individual counterfactual probabilities with model based predictions (Direct Standardization). \emph{Augmented Inverse Probability of Treatment Weighting} (AIPTW) combines both approaches in one estimator, by using a model for the treatment assignment mechanism, a model for the outcome mechanism and, when right-censoring is present, an (optional) model for the censoring mechanism \parencite{Robins1992, Hubbard2000, Ozenne2020}. If right-censoring is completely random or the censoring mechanism model is correct and \emph{either} of the treatment or outcome models is also correctly specified, the resulting estimates are consistent. It therefore gives the researcher "two chances to get it right" \parencite[p. 167]{Hernan2019}.
\par
Multiple slightly different versions of this estimator have been proposed for right-censored data \parencite{Wu2023}. The \texttt{adjustedCurves} package includes the version introduced by \textcite{Ozenne2020}, which is currently only defined for binary treatments, but also works in the presence of competing events. This version technically always requires a model for the censoring mechanism. In practice, a Cox model with an inverted event indicator may be used. Under the completely random right-censoring assumption used througout this paper, however, the censoring model may be replaced by a non-parametric estimate of the cumulative baseline hazard of censoring \parencite{Denz2022}. This is the default used in \texttt{adjustedCurves}.
\par
Due to the form of the estimating equation, it is possible that this method produces estimates that are smaller than 0 or larger than 1. It is also possible that the resulting survival curve is not non-increasing, which means that it may increase again at some point in time. However, theoretical results \parencite{Westling2020} and a simulation study \parencite{Denz2022} indicate that simple post-estimation correction strategies are sufficient to correct the estimates. This correction method consists of truncating out-of-bounds estimates at 0 or 1 and performing isotonic regression afterwards. In the presented package, this estimator can be used by setting \texttt{method="aiptw"} in the \texttt{adjustedsurv()} function. Internally, it relies on the \texttt{ate()} function from the \texttt{riskRegression} package \parencite{Gerds2021}. The post-estimation correction methods are also directly implemented in the \texttt{adjustedsurv()} function and the associated \texttt{plot()} method.

\subsection{Pseudo-Values (PV)} \label{sec::methods_pseudo}

Pseudo-Values (PV) are a clever technique to bypass the need for estimators that can handle right-censoring. The general idea is that one could use standard statistical techniques, such as generalized linear models, for the analysis of survival data, if there was no censoring \parencite{Andersen2017, Andersen2010}. The procedure is straightforward for the estimation of $\hat{S}_{z}(t)$. First, calculate PVs for each individual $i$ at a range of points in time $t$, defined as:

\begin{equation}
	\hat{\theta}_i(t) = n \hat{\theta}(t) - (n - 1) \hat{\theta}^{-i}(t),
\end{equation}

where $\hat{\theta}$ is a non-parametric estimator of the un-adjusted function of interest and $\hat{\theta}^{-1}$ is that same estimator applied to the sample size of size $n$ with the $i$-th observation removed. The simple Kaplan-Meier estimator can be used as $\hat{\theta}$ if right-censoring is random, and an inverse probability of right-censoring weighted Kaplan-Meier estimator may be used if right-censoring depends on observed covariates \parencite{Overgaard2019}.
\par
These PVs can be used directly to perform IPTW (\texttt{method="iptw\_pseudo"}), \\ g-computation (\texttt{method="direct\_pseudo"}) and AIPTW (\texttt{method="aiptw\_pseudo"}). To get an IPTW estimate one can simply take a weighted average of the PVs stratified by the observed treatment status $Z$. The weights can be calculated exactly as described in section~\ref{sec::methods_iptw} \parencite{Andersen2017}. Similarly, by using the PVs as response variable in a generalized estimating equation (GEE) model, one can obtain estimates of the target estimand using g-computation as described in section~\ref{sec::methods_direct} \parencite{Klein2008, Overgaard2017}. Combining both approaches, AIPTW estimates can be calculated \parencite{Wang2018}.
\par
Similar to the AIPTW method, PV based methods may also produce estimates that are smaller than 0 or larger than 1 or are non-mononotonic. The same post-estimation correction methods may be used here \parencite{Denz2022}. Internally our package relies on the \texttt{prodlim} package \parencite{Gerds2019} to obtain the necessary PV and on the \texttt{geepack} package \parencite{Halekoh2006} to calculate the GEE model where necessary.

\subsection{Instrumental Variable based Estimation}

All previously mentioned methods assume that a sufficient adjustment set of confounders has been identified and can be adjusted for. In contrast, instrumental variable based methods allow the estimation of causal effects even in the presence of unmeasured confounders, under certain conditions \parencite{Greenland2000, Lee2023}. An instrumental variable is a variable that fulfills three conditions: (1) it is independent of all unmeasured confounders, (2) it is associated with the treatment variable of interest and (3) it is not a direct cause of the outcome of interest \parencite{Greenland2000}. One method to utilize this kind of variable to estimate causal effects is two-stage least squares regression \parencite{James1978}. Roughly, a linear regression model is esimated, using the treatment of interest as the dependent variable and the instrumental variable as an independent variable. The predictions for the treatment of interest given the first step model and the observed values of the instrumental variable are then used in a second model as a substitute for the actual treatment status. See \textcite{James1978} for more details.
\par
\textcite{MartinezCamblor2020} have recently extended this method to estimate adjusted survival curves. Instead of a linear model, they propose the use of a Cox proportional hazards model with an individual frailty term in the second step. Through the use of the frailty term, they circumvent issues of non-collapsibility of the Cox model. For this reason they refer to their method as \emph{two stage residual inclusion method with a frailty term} (2SRI-F). Importantly, the causal effect estimated by this approach is not, in general, equal to the one defined in equation~\ref{eq::target_estimand}. Rather, under the usual instrumental variable assumptions, it can be interpreted as a local average treatment effect among individuals whose treatment status is affected by the instrumental variable, commonly referred to as compliers \parencite{Imbens1994}.
\par
This method is available in the \texttt{adjustedCurves} package by using \texttt{method="iv\_2SRIF"}. Other related methods through the use of instrumental variables have been proposed by \textcite{Lee2023}, but are currently not implemented in this package. The reason is that it is unclear whether their work generalizes to estimation of the underlying survival curves, as the authors were focussed only on the difference in survival probabilities.

\subsection{Proximal Causal Inference}

Similar to the instrumental variable based method, proximal causal inference based methods allow estimation of counterfactual survival curves in the presence of unmeasured confounding, given specific conditions. The general method was first introduced by \textcite{Miao2018} and later extended to allow time-to-event outcomes by \textcite{Ying2022} and \textcite{Li2025}. Proximal causal inference is designed for settings in which some confounders of the treatment–outcome relationship may be unmeasured, but proxy variables for these unmeasured confounders are available. The key idea is to leverage two types of proxies: treatment-inducing proxies, which are potential causes of the treatment that are only related to the outcome through measured variables and a particular unmeasured confounder $U$, and outcome-inducing proxies, which are potential causes of the outcome, but are only related to the treatment through measured variables and $U$. Under appropriate identifiability assumptions, these proxies can be used to recover causal effects even when standard exchangeability does not hold with respect to the observed covariates alone \parencite{Zivich2023}. More details are given in the cited literature.
\par
Different types of proximal causal inference based methods exists. The \texttt{adjustedCurves} package supports two such methods through the \texttt{adjustedsurv()} function: proximal IPTW (\texttt{method="prox\_iptw"}) and proximal AIPTW (\texttt{method="prox\_aiptw"}). Both methods use particular parametric models for the estimation, that were introduced by \textcite{Ying2022}. The code used was directly adapted from the same original publication, with permission by the authors.

\subsection{Choosing an Estimator}

The large amount of existing estimators naturally raises the question, which one should be used in which situation. This is largely a practical question, as all estimators, with the exception of the instrumental variable based estimator, target the same quantity as defined in equation~\ref{eq::target_estimand}. The choice highly depends on both the features of the observed data, the associated study question, and on which assumptions users are willing to make.
\par
The former is related to practical concerns such as the number of levels in $Z$ and the sample size. For example, if $Z$ has more than two levels, multiple methods (\texttt{method \%in\% c("matching", "emp\_lik", "aiptw", "iv\_2SRIF", "prox\_iptw", "prox\_aiptw")}) \\ may not be used. Similarly, some methods are compuationally very expensive and thus may take a long time to run with large sample sizes. One such example is \texttt{method = "aiptw"}, as implemented in the presented package. If many points of time are of interest, the sample size is large and approximate standard error calculations are needed, it may not be computationally feasible to use this method. Simpler methods, such as \texttt{method = "iptw\_km"} or \texttt{method = "direct"} may be preferable in this case. Conversely, a previous simulation study has shown that some methods, such as \texttt{method = "direct"}, do not work well with small sample sizes \parencite{Denz2022}.
\par
Another important point to consider is the adjustment procedure itself. In some cases it might be easier to model the treatment assignment process than it is to model the outcome process or vice versa. If possible, we recommend using doubly-robust methods, such as \texttt{method="aiptw"} or \texttt{method="aiptw\_pseudo"}, due to their robustness to model misspecification. If usage of these methods is not possible, but users are confident that either the treatment or outcome model is well-specified, the associated simpler methods may be used. In the special case where all relevant confounders are binary or categorical and the sample size in each stratum is sufficiently large, usage of non-parametric stratification methods such as \texttt{method="strat\_amato"} may be the best choice, to avoid modelling entirely.
\par
Apart from these practical issues, users should also carefully consider which assumptions are most suitable. For example, if clinical or scientific knowledge suggests that an important confounder is unmeasured, most methods implemented in this package may not produce unbiased estimates of the target quantity. In this case, users may check whether suitable proxies for the application of proximal causal inference based methods \parencite{Miao2018} or a suitable instrument for the application of \texttt{method="iv\_2SRIF"} \parencite{MartinezCamblor2020} exist. Similarly, users should carefully asses whether censoring can be assumed to be independent of covariates. If this is not the case, explicit adjustment for dependent censoring is needed, which is not supported by all methods. An overview of all of the points mentioned here is given in tables~\ref{tab::method_comparison_1}--\ref{tab::method_comparison_4}.

\begin{table}[htbp]
	\centering
	\small
	\setlength{\tabcolsep}{3pt}
	\renewcommand{\arraystretch}{1.2}
	
	\caption{An overview of the implemented functionalities of each method in the \texttt{adjustedsurv()} function (part 1).}
	\label{tab::method_comparison_1}
	
	\begin{tabular}{@{}LCCCC@{}}
		\toprule
		&
		\texttt{"direct"}
		&
		\texttt{"direct\_pseudo"}
		&
		\texttt{"iptw\_km"}
		&
		\texttt{"iptw\_cox"}
		\\
		\midrule
		
		\textbf{Unmeasured Confounding}
		& no & no & no & no \\
		
		\textbf{Categorical Treatments}
		& yes & yes & yes & yes \\
		
		\textbf{Continuous Confounders}
		& yes & yes & yes & yes \\
		
		\textbf{Approximate SE}
		& yes & no & yes & no \\
		
		\textbf{In Bounds}
		& yes & yes & yes & yes \\
		
		\textbf{Non-increasing}
		& yes & no & yes & yes \\
		
		\textbf{Doubly Robust}
		& no & no & no & no \\
		
		\textbf{Dependent Censoring}
		& no & yes & no & no \\
		
		\textbf{Adjustment Type}
		& outcome & outcome & treatment & treatment \\
		
		\textbf{Nonparametric}
		& no & no & depends & depends \\
		
		\textbf{Computation Speed}
		& + & -\,- & ++ & ++ \\
		
		\textbf{Dependencies}
		& \texttt{riskRegression}
		& \texttt{geepack}, \texttt{prodlim}
		& -
		& -
		\\
		
		\bottomrule
	\end{tabular}
\end{table}

\begin{table}[htbp]
	\centering
	\small
	\setlength{\tabcolsep}{3pt}
	\renewcommand{\arraystretch}{1.2}
	
	\caption{An overview of the implemented functionalities of each method in the \texttt{adjustedsurv()} function (part 2).}
	\label{tab::method_comparison_2}
	
	\begin{tabular}{@{}LCCCC@{}}
		\toprule
		&
		\texttt{"iptw\_pseudo"}
		&
		\texttt{"matching"}
		&
		\texttt{"emp\_lik"}
		&
		\texttt{"aiptw"}
		\\
		\midrule
		
		\textbf{Unmeasured Confounding}
		& no & no & no & no \\
		
		\textbf{Categorical Treatments}
		& yes & no & no & no \\
		
		\textbf{Continuous Confounders}
		& yes & yes & yes & yes \\
		
		\textbf{Approximate SE}
		& yes & no & no & yes \\
		
		\textbf{In Bounds}
		& no & yes & yes & no \\
		
		\textbf{Non-increasing}
		& no & yes & yes & no \\
		
		\textbf{Doubly Robust}
		& no & no & no & yes \\
		
		\textbf{Dependent Censoring}
		& yes & no & no & yes \\
		
		\textbf{Adjustment Type}
		& treatment & treatment & treatment & both \\
		
		\textbf{Nonparametric}
		& depends & depends & yes & no \\
		
		\textbf{Computation Speed}
		& - & - & + & - \\
		
		\textbf{Dependencies}
		& \texttt{prodlim}
		& \texttt{Matching}
		& \texttt{MASS}
		& \texttt{riskRegression}
		\\
		
		\bottomrule
	\end{tabular}
\end{table}

\begin{table}[htbp]
	\centering
	\small
	\setlength{\tabcolsep}{3pt}
	\renewcommand{\arraystretch}{1.2}
	
	\caption{An overview of the implemented functionalities of each method in the \texttt{adjustedsurv()} function (part 3).}
	\label{tab::method_comparison_3}
	
	\begin{tabular}{@{}LCCCC@{}}
		\toprule
		&
		\texttt{"aiptw\_pseudo"}
		&
		\texttt{"strat\_amato"}
		&
		\texttt{"strat\_nieto"}
		&
		\texttt{"strat\_cupples"}
		\\
		\midrule
		
		\textbf{Unmeasured Confounding}
		& no & no & no & no \\
		
		\textbf{Categorical Treatments}
		& yes & yes & yes & yes \\
		
		\textbf{Continuous Confounders}
		& yes & no & no & no \\
		
		\textbf{Approximate SE}
		& yes & no & yes & no \\
		
		\textbf{In Bounds}
		& no & yes & yes & yes \\
		
		\textbf{Non-increasing}
		& no & yes & yes & yes \\
		
		\textbf{Doubly Robust}
		& yes & no & no & no \\
		
		\textbf{Dependent Censoring}
		& yes & no & no & no \\
		
		\textbf{Adjustment Type}
		& both & - & - & - \\
		
		\textbf{Nonparametric}
		& no & yes & yes & yes \\
		
		\textbf{Computation Speed}
		& -\,- & +++ & +++ & +++ \\
		
		\textbf{Dependencies}
		& \texttt{geepack}, \texttt{prodlim}
		& -
		& -
		& -
		\\
		
		\bottomrule
	\end{tabular}
\end{table}

\begin{table}[htbp]
	\centering
	\small
	\setlength{\tabcolsep}{3pt}
	\renewcommand{\arraystretch}{1.2}
	
	\caption{An overview of the implemented functionalities of each method in the \texttt{adjustedsurv()} function (part 4).}
	\label{tab::method_comparison_4}
	
	\begin{tabular}{@{}LCCCC@{}}
		\toprule
		&
		\texttt{"iv\_2SRIF"}
		&
		\texttt{"prox\_iptw"}
		&
		\texttt{"prox\_aiptw"}
		&
		\texttt{"km"}
		\\
		\midrule
		
		\textbf{Unmeasured Confounding}
		& yes & yes & yes & no \\
		
		\textbf{Categorical Treatments}
		& no & no & no & yes \\
		
		\textbf{Continuous Confounders}
		& yes & yes & yes & no \\
		
		\textbf{Approximate SE}
		& no & yes & yes & yes \\
		
		\textbf{In Bounds}
		& yes & no & no & yes \\
		
		\textbf{Non-increasing}
		& yes & no & no & yes \\
		
		\textbf{Doubly Robust}
		& no & no & yes & no \\
		
		\textbf{Dependent Censoring}
		& no & no & no & no \\
		
		\textbf{Adjustment Type}
		& - & treatment & both & none \\
		
		\textbf{Nonparametric}
		& no & no & no & yes \\
		
		\textbf{Computation Speed}
		& + & -\,- & -\,- & +++ \\
		
		\textbf{Dependencies}
		& -
		& \texttt{numDeriv}
		& \texttt{numDeriv}
		& -
		\\
		
		\bottomrule
	\end{tabular}
\end{table}

\clearpage
\newpage

\section{Implementation}

\subsection{Existing Software}

Before describing the workflow and implementation in the proposed package, we review already existing implementations and related software packages. Although this article is mainly focused on \textbf{R} software, there are also some implementations in the \textbf{Stata} and \textbf{SAS} programming environments. The \texttt{stpm2\_standsurv} command can be used to generate g-computation estimates in \textbf{Stata}, after a survival model has been fitted \parencite{Lambert2017, Lambert2018}. Similarly, there are multiple \textbf{SAS} routines to calculate g-computation based estimates of the target-estimand \parencite{Zhang2007, Zhang2011, Hu2021}. Additionally, some \textbf{SAS} code for the estimation of IPTW adjusted survival curves is available as supplemental material in the article of \textcite{Cole2004} and some \textbf{SAS} code is also available for the estimation of AIPTW adjusted survival curves in the supplemental material of \textcite{Zhang2012}. Lastly, the calculation of Pseudo-Values is also implemented in \textbf{SAS} \parencite{Klein2008}, but the actual adjustment methods are, to the best of our knowledge, not.
\par
In the \textbf{R} programming language there are multiple packages implementing some of the available methods. The popular \texttt{survminer} package \parencite{Kassambara2021}, contains the \texttt{ggadjustedcurves()} function, which implements g-computation (using \\ \texttt{method="conditional"}) to adjust survival curves for confounders using a previously fit Cox model. Another implementation of g-computation for the estimation of confounder-adjusted survival curves can be found in the \texttt{RISCA} package \parencite{Foucher2020}, which also implements one form of IPTW with an associated adjusted Log-Rank test. An implementation of the EL method is available in the \texttt{adjKMtest} package \parencite{Wang2019} available on github (\url{https://github.com/kimihua1995/adjKMtest}), while the \texttt{survtmle} package \parencite{Benkeser2019} implements the targeted maximum likelihood based estimator of the target estimand for discrete time-to-event data. Most notably, the \texttt{riskRegression} package \parencite{Gerds2021} contains the \texttt{ate()} function \parencite{Ozenne2020}, which can be used to calculate g-computation, IPTW and AIPTW estimates of the adjusted survival curves and adjusted cause-specific CIFs.
\par
A summary of the properties of all existing \textbf{R} packages on the topic that we know of is presented in table~\ref{tab::r_packages}. These packages have a vastly different syntax, making it time-consuming to try out multiple methods. Additionally, all available implementations in \textbf{R}, \textbf{Stata} and \textbf{SAS} only offer limited functionality. Many do not allow the user to calculate confidence intervals, others only allow binary treatments or have no plot function. There is also no publicly available software which directly implements the stratification and PV based methods. The \textbf{R} package presented in this article tries to fix those shortcomings.

\begin{table}[!htb]
	\centering
	\caption{An overview of all \textbf{R} packages implementing one or more of the discussed methods for confounder-adjustment. Methods dealing strictly with other scenarios, such as estimation in the presence of time-dependent confounding, are not included in this table.}
	\label{tab::r_packages}
	\par\medskip\smallskip
	\begin{tabular}{p{2.8cm} p{3.1cm} p{1.2cm} p{1.9cm} p{1.6cm} p{1.6cm} p{1.3cm}}
		\toprule
		Name & Authors & On CRAN & Methods & Summary Statistics & Confidence Intervals & Plot Function \\
		\midrule
		\texttt{survminer} & \textcite{Kassambara2021} & Yes & G-Comp. & No & No & Yes \\
		\texttt{stdReg} & \textcite{Sjoelander2016} & Yes & G-Comp. & No & Yes & Yes \\
		\texttt{survey} & \textcite{Lumley2004} & Yes & IPTW & No & Yes & Yes \\
		\texttt{RISCA} & \textcite{Foucher2020} & Yes & G-Comp., IPTW & No & Partially & Yes \\
		\texttt{riskRegression} & \textcite{Gerds2021} & Yes & G-Comp., IPTW, AIPTW & Yes & Yes & No \\
		\texttt{survtmle} & \textcite{Benkeser2019} & No & TMLE & Yes & Yes & No \\
		\texttt{adjKMtest} & \textcite{Wang2019} & No & EL, IPTW & No & Yes & No \\
		\texttt{MOSS} & \textcite{Cai2020} & No & TMLE & Yes & Yes & Yes \\
		\texttt{survdr} & \textcite{Diaz2019} & No & TMLE & No & No & No \\
		\texttt{CFsurvival} & \textcite{Westling2021} & No & AIPTW & No & Yes & No \\
		\texttt{OW\_Survival} & \textcite{Cheng2022} & No & IPTW & No & Yes & Yes \\
		\texttt{AdjKM.CIF} & \textcite{Cao2022} & No & G-Comp. & Yes & Yes & Yes \\
		\texttt{concrete} & \textcite{Chen2023} & No & TMLE & Yes & Yes & Yes \\
		\texttt{adjSURVCI} & \textcite{Khanal2023} & Yes & G-Comp. & No & Yes & No \\
		\texttt{causalBETA} & \textcite{Ji2023} & No & G-Comp. & Yes & Yes & Yes \\
		\texttt{pci2s} & \textcite{Li2025} & No & Prox. G-Comp & No & No & No \\
		\bottomrule
	\end{tabular}
\end{table}

\FloatBarrier

\clearpage

\subsection{Implementation in the adjustedCurves package}

The main function of the \texttt{adjustedCurves} package is the \texttt{adjustedsurv()} function, which implements the estimation of confounder-adjusted survival curves using all of the methods mentioned above. It has the following syntax:
\par

\begin{lstlisting}[style=mystyle]
adjustedsurv(data, variable, ev_time, event, method,
             conf_int=FALSE, conf_level=0.95, times=NULL,
             bootstrap=FALSE, n_boot=500,
             n_cores=1, na.action=options()$na.action,
             clean_data=TRUE, iso_reg=FALSE,
             force_bounds=FALSE, mi_extrapolation=FALSE,
             ...)
\end{lstlisting}

\par
The required arguments are:

\begin{itemize}
	\item \texttt{data}: The data set including all needed variables, or a \texttt{mids} object containing multiply imputed data created using the \texttt{mice} package \parencite{vanBuuren2011}.
	\item \texttt{variable}: The name of the treatment variable ($Z$).
	\item \texttt{ev\_time}: The name of the variable specifying the observed failure times ($T_{obs}$).
	\item \texttt{event}: The name of the variable indicating the event status ($D$).
	\item \texttt{method}: A character string specifying the adjustment method to use.
\end{itemize}

As the description of the \texttt{data} argument suggests, multiply imputation to deal with missing values \parencite{vanBuuren2011} is directly supported for all methods. Depending on the \texttt{method} used, other arguments might be required. For example when using \texttt{method="direct"}, the user should also supply a suitable model to make conditional survival probability predictions to the \texttt{outcome\_model} argument. All method-specific required arguments are listed on the documentation page of the respective method, similar to the design of the \texttt{WeightIt} \parencite{Greifer2021} and \texttt{MatchIt} \parencite{Ho2011} packages. In case one of these arguments is missing, the \texttt{adjustedsurv()} function will inform the user by printing helpful error messages.
\par
Some further optional arguments to the function are:

\begin{itemize}
	\item \texttt{conf\_int}: Whether to calculate approximate point-wise confidence intervals. Not available for all methods. Availability is displayed in tables~\ref{tab::method_comparison_1}--\ref{tab::method_comparison_4}.
	\item \texttt{conf\_level}: A number specifying the confidence level of asymptotic and/or bootstrap confidence intervals.
	\item \texttt{times}: Specific points in time for which the adjusted survival probability should be estimated.
	\item \texttt{bootstrap}: Whether to perform bootstrapping or not.
	\item \texttt{n\_boot}: The number of bootstrap replications to perform if \texttt{bootstrap=TRUE}.
	\item \texttt{n\_cores}: The number of processor cores to use for bootstrapping.
	\item \texttt{na.action}: How to handle missing values.
	\item \texttt{clean\_data}: Whether to only keep relevant variables before applying the specified \texttt{na.action}.
	\item \texttt{iso\_reg}: Whether to apply isotonic regression to the resulting survival curves to ensure that they are non-increasing. This may not be the case when using \\ \texttt{method="direct\_pseudo"}, \texttt{method="iptw\_pseudo"}, \texttt{method="aiptw\_pseudo"} or \\ \texttt{method="aiptw"}. Isotonic regression fixes this problem without introducing any asymptotic bias \parencite{Westling2020}.
	\item \texttt{force\_bounds}: Whether to truncate survival probabilities larger than 1 or smaller than 0, which also may occur for some methods as discussed above.
	\item \texttt{mi\_extrapolation}: Whether to allow potential extrapolation of survival curves due to imputed values. Only relevant when using multiply imputed data.
	\item \texttt{...}: Additional method-specific optional arguments.
\end{itemize}

This function returns an \texttt{adjustedsurv} object, which includes a \texttt{data.frame} of the treatment-specific causal survival curves and other useful objects, such as the bootstrapped estimates if bootstrapping was performed. To plot these estimates, the \texttt{plot()} function can be used. Since this function is designed to create highly customizable graphics, it has a lot of optional arguments. We will only list the most important ones here:

\par

\begin{lstlisting}[style=mystyle]
plot(adjsurv, conf_int=FALSE, cif=FALSE, color=TRUE,
     linetype=FALSE, facet=FALSE, median_surv_lines=FALSE,
     censoring_ind="none", risk_table=FALSE, ...)
\end{lstlisting}

\par

Some of these arguments are:

\begin{itemize}
	\item \texttt{adjsurv}: An object created by the \texttt{adjustedsurv()} function.
	\item \texttt{conf\_int}: Whether to draw the confidence intervals or not.
	\item \texttt{cif}: If set to \texttt{TRUE}, the CIF is displayed instead of the survival curve.
	\item \texttt{color}, \texttt{linetype}, \texttt{facet}: How to differentiate between the treatment groups.
	\item \texttt{median\_surv\_lines}: Automatically add indicator lines for the median survival time.
	\item \texttt{censoring\_ind}: Whether and how to draw indicators showing when censored observations occurred.
	\item \texttt{risk\_table}: Whether to add a table including numbers at risk or numbers of censored observations in the plot.
\end{itemize}

The function returns a \texttt{ggplot} object, which can be further processed using standard \texttt{ggplot2} syntax. Custom colors, sizes, linetypes and many more things can be defined using other available arguments. A full list can be found on the documentation page.

\section{Illustrative Example}

To illustrate the functionality of the proposed package, we will use data from the Rotterdam tumor bank, which includes 2982 primary breast cancer patients \parencite{Foekens2000}. The data set is publicly available as part of the \texttt{survival} \textbf{R} package \parencite{Therneau2021}, allowing anyone to replicate the results presented here. It includes information on the time until death or censoring as well as additional information about the patients, such as age, tumor size, and whether the patient received a hormonal treatment. Hormonal therapy is a widely used treatment for breast cancer. Multiple randomized controlled trials have shown that it reduces the probability of recurrence and death in primary breast cancer patients, particularly when used in combination with chemotherapy \parencite{Hortobagyi1998, Abdulkareem2012}.
\par
For exemplary purposes, we will assume that we are interested in assessing the total causal effect of hormonal therapy on all-cause survival using the Rotterdam tumor bank data. However, the analysis carried out here is done strictly for illustrative purposes. Our goal is not to generate new knowledge about this treatment strategy, we merely use it as an example to show why confounder-adjusted survival curves are an important tool and how to use them. Clinical results obtained using this data set can be found in the original article by \textcite{Foekens2000}.
\par
First, we load all packages that will be required for the analysis performed in this article. Afterwards we estimate and plot simple Kaplan-Meier survival curves for each group of interest. We set the \texttt{conf\_int} argument to \texttt{TRUE} in order to obtain and plot the associated point-wise 95\% confidence interval of the survival curve. Note that the treatment variable always needs to be a \texttt{factor} variable in order to use the \texttt{adjustedsurv()} function. We use the following code:

\par

\begin{lstlisting}[style=mystyle]
library("survival")
library("pammtools")
library("ggplot2")
library("riskRegression")
library("adjustedCurves")
	
rotterdam$hormon <- factor(rotterdam$hormon,
                           labels=c("No", "Yes"))
	
km <- adjustedsurv(data=rotterdam,
                   variable="hormon",
                   ev_time="dtime",
                   event="death",
                   method="km",
                   conf_int=TRUE)
	
plot(km, conf_int=TRUE, legend.title="Hormonal\nTherapy",
     gg_theme=theme_minimal(), x_breaks=seq(0, 7000, 1000),
     risk_table=TRUE, risk_table_stratify=TRUE)
\end{lstlisting}

\par

\begin{center}
	\includegraphics[width=\linewidth]{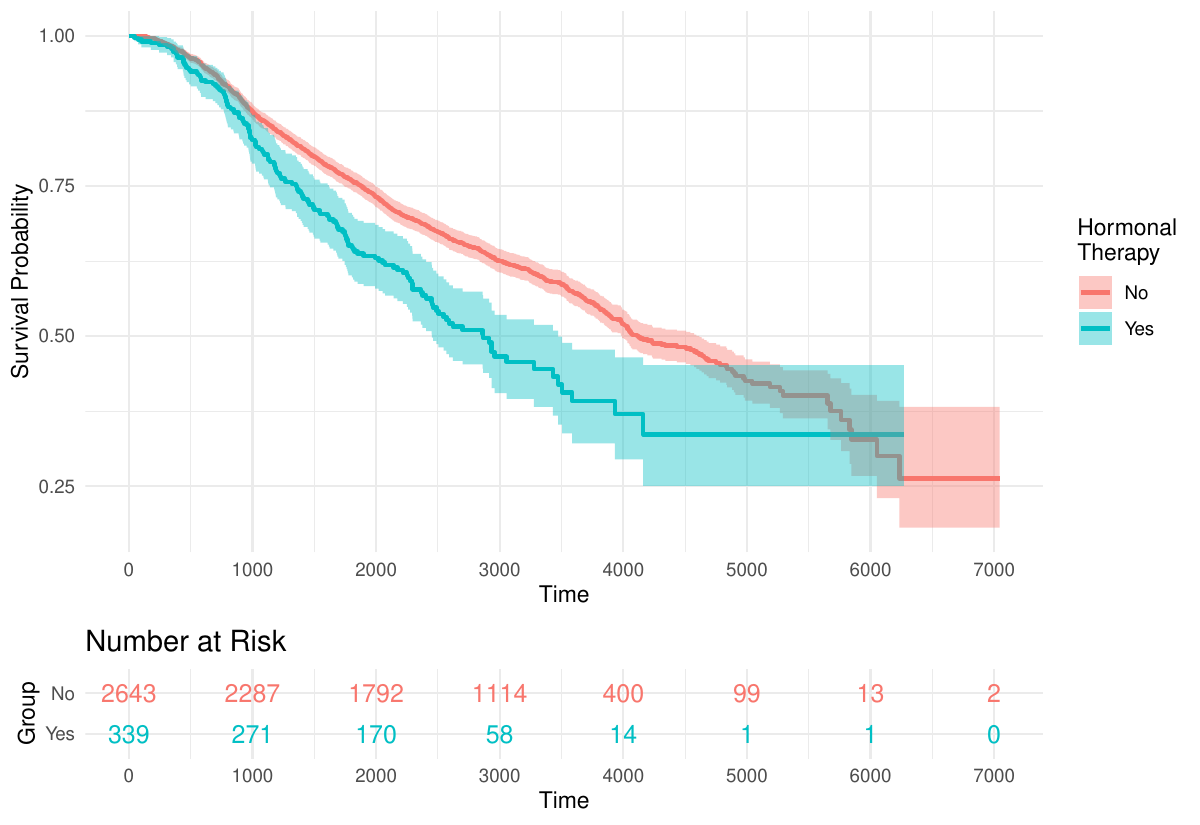}
\end{center}

It is important to note that the confidence intervals only have point-wise 95\% coverage, as they do not represent simultaneous confidence bands \parencite{Sachs2022}. When using such plots in real studies, the nature of the confidence intervals should be stated directly in the figure legend to avoid confusion. In addition to the confidence intervals, we also added the numbers at risk over time for each treatment group to the plot by setting the \texttt{risk\_table} and \texttt{risk\_table\_stratify} arguments in \texttt{plot.adjustedsurv()} to \texttt{TRUE}. Furthermore, we changed the \texttt{ggplot2} theme and the breaks of the $x$-axis for more visually appealing results. One can clearly see a big difference in the survival probability between patients who received a hormonal therapy and patients who did not receive it. Patients with hormonal therapy seem to have a drastically lower survival probability than other patients over most of the considered time. In large parts of the plot (between approximately 1000 and 4500 days after study begin), the 95\% confidence intervals do not overlap, indicating that this point-wise difference is statistically significant as well.
\par
This counter intuitive result is, however, not due to any actual harmful effects of the hormonal therapy. The real reason for the large difference between these two treatment groups is due to confounding. Patients with a worse prognosis tend to receive hormonal therapy more often than patients with a better prognosis \parencite{Hortobagyi1998, Abdulkareem2012}. The standard Kaplan-Meier curves shown above therefore do not give an unbiased estimate of the true counterfactual survival curves. Instead, we need to use one of the methods discussed above to adjust for the relevant confounders.
\par
We assume that the age of the patient at surgery (\texttt{age}), the number of positive lymph nodes (\texttt{nodes}), the size of the tumor \texttt{size}, the differentiation grade (\texttt{grade}) and menopausal status (\texttt{meno}) form a sufficient adjustment set for the treatment-outcome relationship. In other words, we assume that it is sufficient to adjust for these confounders in order to get an unbiased estimate of the total causal effect of the hormonal therapy on all-cause survival. We further assume that the causal identifiability assumptions stated in section~\ref{sec::target_estimand} hold for simplicity. In a real data analysis, development of a causal directed acyclic graph to identify a sufficient adjustment set based on expert opinion and the available literature would be recommended.
\par
Below we use stabilized inverse probability of treatment weights to perform the adjustment. First, we fit a logistic regression model with the treatment status as response variable and the relevant confounders as independent variables. Afterwards, we use the \texttt{adjustedsurv()} function with \texttt{method="iptw\_km"} and \texttt{stabilize=TRUE} to obtain the desired estimates. The only difference to creating standard Kaplan-Meier curves is that the propensity score model has to be passed to the \texttt{treatment\_model} argument.

\par

\begin{lstlisting}[style=mystyle]
ps_model <- glm(hormon ~ age + nodes + size + grade + meno,
                data=rotterdam, family="binomial")
	
iptw <- adjustedsurv(data=rotterdam,
                     variable="hormon",
                     ev_time="dtime",
                     event="death",
                     method="iptw_km",
                     treatment_model=ps_model,
                     stabilize=TRUE,
                     conf_int=TRUE)
	
plot(iptw, conf_int=TRUE, legend.title="Hormonal\nTherapy",
     gg_theme=theme_minimal(), x_breaks=seq(0, 7000, 1000),
     risk_table=TRUE, risk_table_stratify=TRUE)
\end{lstlisting}

\par

\begin{center}
	\includegraphics[width=0.9\linewidth]{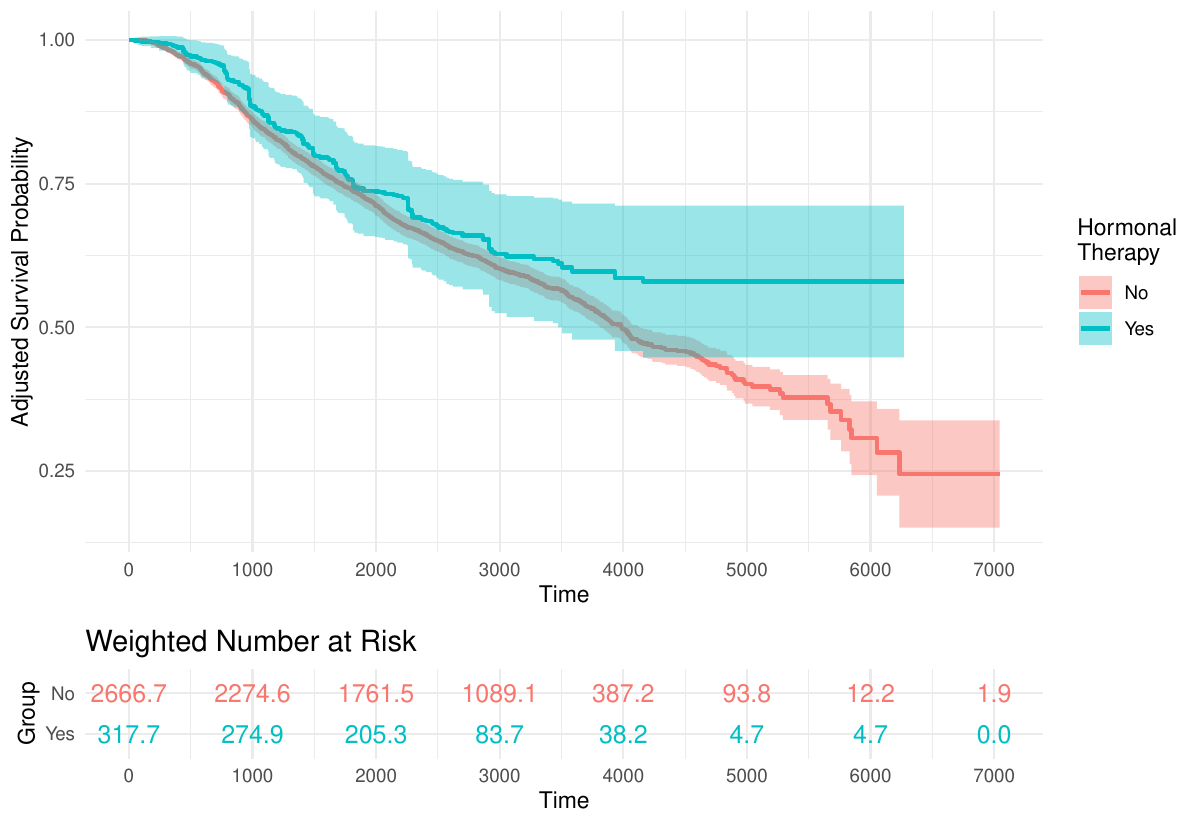}
\end{center}

The resulting curves are much closer to each other than the un-adjusted Kaplan-Meier curves. In fact, there seems to be no discernible difference between the survival curves of the two treatment groups anymore. The curves and their 95\% confidence intervals are overlapping over most of the considered time. Note that since the employed estimator is directly based on a weighted Kaplan-Meier estimator \parencite{Xie2005}, we also choose to display the corresponding numbers at risk. These are not the same crude numbers as shown before, but the weighted ones. Due to the weights, fractional numbers at risk are observed. We could hide these fractions by setting \texttt{risk\_table\_digits=0} to round them to the nearest integer, but choose not to in this case to highlight the difference.
\par
Alternatively, we may use g-computation to obtain the confounder-adjusted survival curves. The workflow is very similar here. Instead of fitting a propensity score model, we now have to fit a model with the time-to-event outcome as a response variable. In this case, we use a simple Cox model to do this, again including the relevant confounders as independent variables in the model. Note that the \texttt{adjustedsurv()} function requires us to use \texttt{x=TRUE} in the \texttt{coxph()} call to work properly. After the model has been fit, we can simply call the \texttt{adjustedsurv()} function using \texttt{method="direct"} by passing the Cox model object to the \texttt{outcome\_model} argument.

\par

\begin{lstlisting}[style=mystyle]
cox_model <- coxph(Surv(dtime, death) ~ hormon + age +
                   nodes + size + grade + meno,
                   data=rotterdam, x=TRUE)
	
direct <- adjustedsurv(data=rotterdam,
                       variable="hormon",
                       ev_time="dtime",
                       event="death",
                       method="direct",
                       outcome_model=cox_model,
                       conf_int=TRUE)
	
plot(direct, conf_int=TRUE, legend.title="Hormonal\nTherapy",
     gg_theme=theme_minimal(), x_breaks=seq(0, 7000, 1000))
\end{lstlisting}

\par

\begin{center}
	\includegraphics[width=1\linewidth]{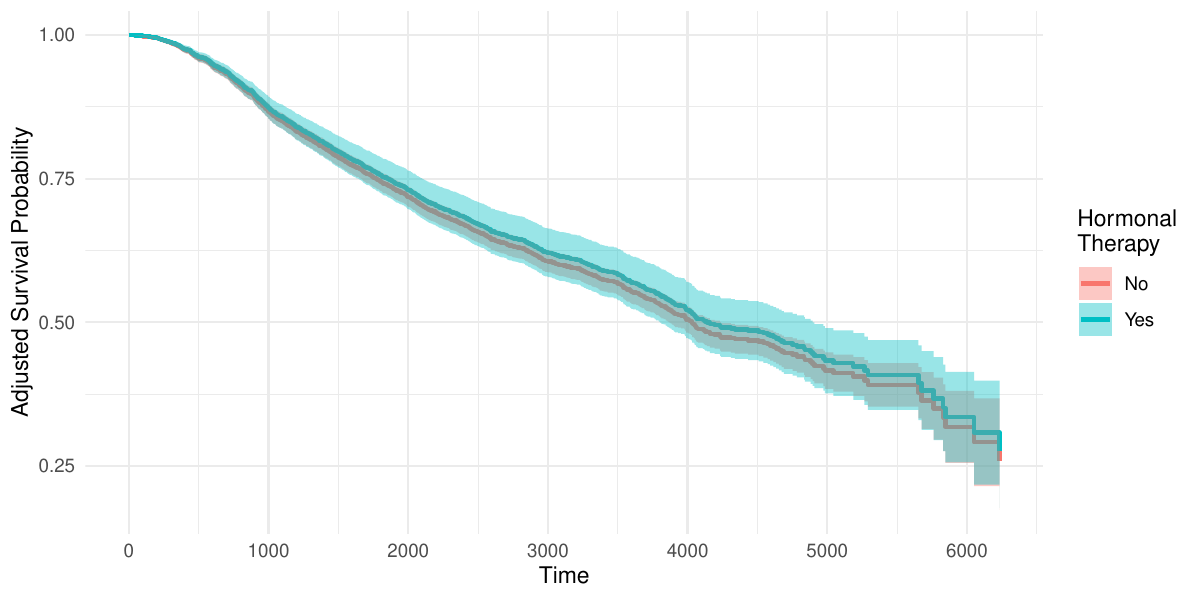}
\end{center}

Contrary to the IPTW case, we did not add numbers at risk to this plot, because the estimation does not directly rely on these values. Adding crude numbers at risk would be possible, but might be confusing because those do not correspond to the shown curves. Nevertheless, the results are very similar to the IPTW adjusted curves. Because g-computation is generally more efficient \parencite{Denz2022} the 95\% confidence intervals are a little bit smaller than before. As discussed in chapter~\ref{sec::methods}, both of these methods require that the models are correctly specified. If we want to relax this assumption, we could use augmented inverse probability of treatment weighting, where only one of the models needs to be correct in order to obtain correct results. In the \texttt{adjustedCurves} package, we can do this using the following code:

\par

\begin{lstlisting}[style=mystyle]
aiptw <- adjustedsurv(data=rotterdam,
                      variable="hormon",
                      ev_time="dtime",
                      event="death",
                      method="aiptw",
                      treatment_model=ps_model,
                      outcome_model=cox_model,
                      bootstrap=TRUE,
                      n_boot=500)
	
plot(aiptw, iso_reg=TRUE, legend.title="Hormonal\nTherapy",
     gg_theme=theme_minimal(), x_breaks=seq(0, 7000, 1000),
     conf_int=TRUE, use_boot=TRUE)
\end{lstlisting}

\par

\begin{center}
	\includegraphics[width=\linewidth]{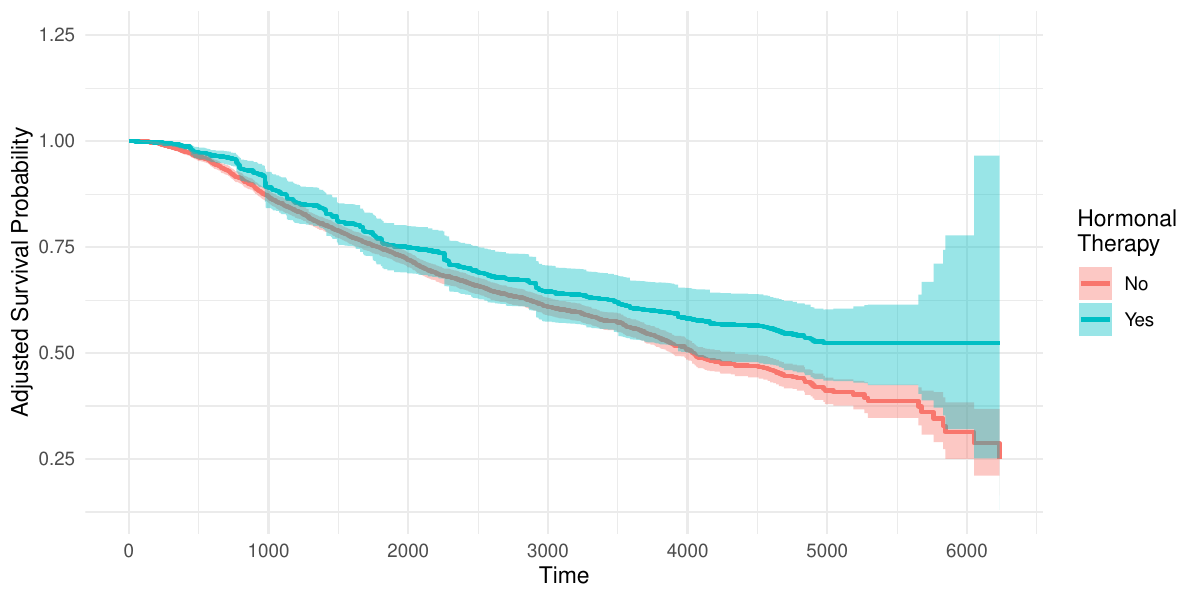}
\end{center}

We simply set the \texttt{method} argument to \texttt{"aiptw"} and pass the propensity score model to the \texttt{treatment\_model} argument and the Cox model to the \texttt{outcome\_model} argument. When plotting the resulting estimates, we set \texttt{iso\_reg=TRUE} to automatically use isotonic regression, which ensures that the survival curves are non-increasing \parencite{Westling2020}. Note that we did not calculate confidence intervals the same way as before here, because the computation is very time extensive. Instead, we used \texttt{bootstrap=TRUE} with \texttt{n\_boot=500} to obtain boostrap-based confidence intervals based on 500 bootstrap samples. In this case, the curves are very similar to the estimates obtained using IPTW and g-computation.
\par
In addition to simply plotting the survival curves, we can compute estimates of the average treatment effect directly based on the estimated survival probabilities to compare the two treatment groups as well. For example, the difference between the survival probabilities after five, six and seven years can be calculated from each of the \texttt{adjustedsurv} objects created above using the \texttt{adjusted\_curve\_diff()} function. Below we demonstrate this using the g-computation based estimates:

\par
\newpage

\begin{lstlisting}[style=mystyle]
adjusted_curve_diff(direct,
                    times=c(365*5, 365*6, 365*7),
                    conf_int=TRUE)
\end{lstlisting}

\par

\begin{framed}
	\begin{verbatim}
		time        diff         se    ci_lower   ci_upper   p_value
		1 1825 -0.01115577 0.01823477 -0.04689527 0.02458373 0.5406798
		2 2190 -0.01267147 0.02022278 -0.05230739 0.02696444 0.5309253
		3 2555 -0.01379431 0.02190894 -0.05673505 0.02914643 0.5289431
	\end{verbatim}
\end{framed}

By setting \texttt{conf\_int=TRUE} again, we also obtain the corresponding confidence intervals of the point-differences and a $p$~value of a simple t-test. In this case, the differences are very small with confidence intervals that include 0 for all considered points in time, indicating that there is no difference in the survival probability after five, six or seven years. To depict this difference directly, the \texttt{plot\_curve\_diff()} function can be used:

\par

\begin{lstlisting}[style=mystyle]
plot_curve_diff(direct, conf_int=TRUE, gg_theme=theme_minimal())
\end{lstlisting}

\par

\begin{center}
	\includegraphics[width=\linewidth]{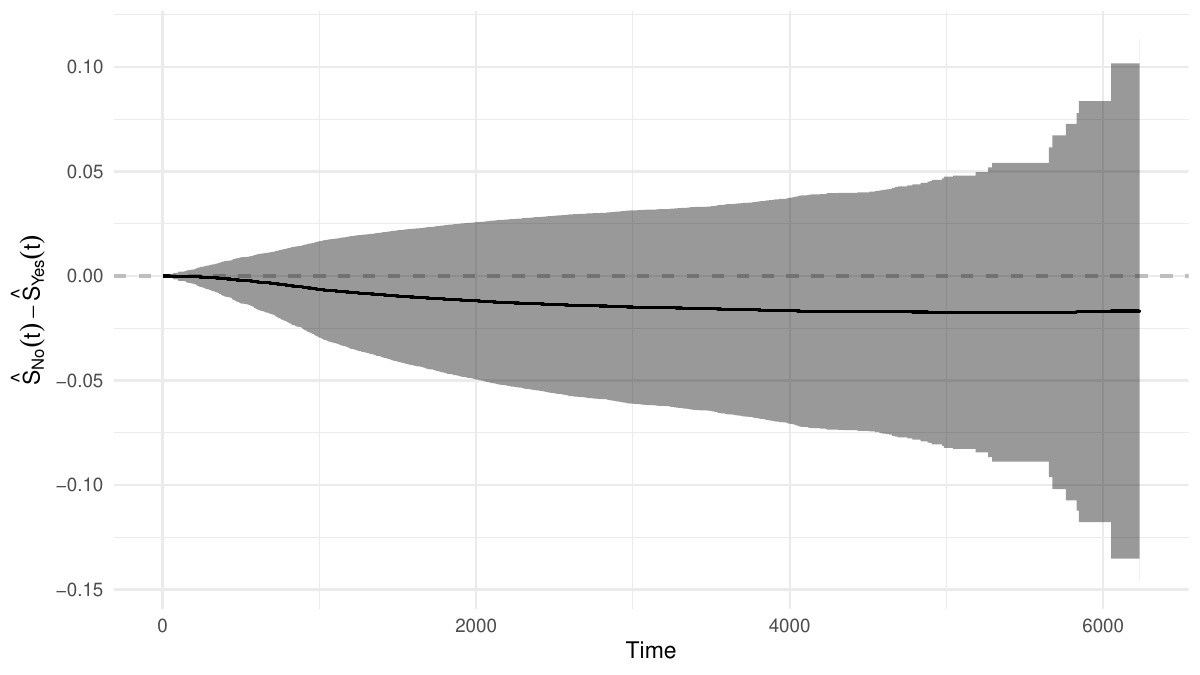}
\end{center}

\par

which clearly shows that, at least for the g-computation based estimates, there is no statistically significant difference for any point in time.

\newpage

The median survival time for each group may also be calculated in a straightforward manner using the \texttt{adjusted\_surv\_quantile()} function:

\par

\begin{lstlisting}[style=mystyle]
adjusted_surv_quantile(direct, p=0.5, conf_int=TRUE)
\end{lstlisting}

\par

\begin{framed}
	\begin{verbatim}
		p group q_surv ci_lower ci_upper
		1 0.5    No   4032     3869     4239
		2 0.5   Yes   4138     3803     4895
	\end{verbatim}
\end{framed}

\par

This function calculates general survival quantiles. By using \texttt{p=0.5}, the median survival time is calculated. Again, no substantial differences can be seen. The adjusted restricted mean survival time may also be calculated by integrating the estimated counterfactual survival curves up to a specified point in time using the \texttt{adjusted\_rmst()} function. Here, we use five years as a reference value:

\par

\begin{lstlisting}[style=mystyle]
adjusted_rmst(direct, to=365*5)
\end{lstlisting}

\par

\begin{framed}
	\begin{verbatim}
		to group     rmst
		1 1825    No 1615.429
		2 1825   Yes 1625.164
	\end{verbatim}
\end{framed}

\par

Again, the two values are nearly the same. If we wanted to obtain confidence intervals for these estimates, we would have to use bootstrapping (\texttt{bootstrap=TRUE}) in the original \texttt{adjustedsurv()} function call, because approximate or exact equations for this method have not been proposed yet in the literature.

\section{Discussion} \label{sec:summary}

In this paper we presented the \texttt{adjustedCurves} \textbf{R} package for the estimation of confounder-adjusted survival curves and average treatment effects. Contrary to other available packages, our package offers wrappers and implementations of most available methods using one consistent framework. It has native support for the calculation of confidence intervals using both approximate equations (where available) and the computationally more expensive bootstrapping. It also directly supports multiple imputation to naturally handle missing data, which is an important concern when considering multiple covariates. As shown in the examples, the included \texttt{plot()} functions can be used effectively to create publication ready graphical displays of the adjusted curves, potentially including risk tables. Additionally, it can be used to directly compare groups using associated statistics such as differences in probabilities or the restricted mean survival time.
\par
Although the need for confounder-adjustment when analyzing observational data is well known, adjustment-methods for survival curves are underutilized in practice. We hope that this package facilitates the usage of these methods by making it easier for researchers to apply them. The different types of survival curve based marginal treatment effects were implemented for the same reason. Multiple researchers have advocated for using these measures instead of the conditional hazard ratio \parencite{Royston2013, Hasegawa2020}, because they are arguable easier to interpret and thus make communication of results easier. In contrast to the hazard ratio, they also allow a causal interpretation of the results \parencite{Aalen2015}.
\par
To keep the article at a reasonable length, we did not discuss all features of the package in this article. One important additional feature is the support for data with competing events, in which the occurrence of one type of event may preclude the occurrence of another type of event \parencite{Satagopan2004}. For example, when the main event of interest is the time to the development of a disease, such as cancer, death unrelated to cancer may be considered a competing event. In such situations, the estimation of survival probabilities specific to a cause is generally not possible, crude or adjusted. However, it is possible to estimate cause-specific cumulative incidences instead \parencite{Scrucca2007}. In the proposed package, the estimation of confounder-adjusted cause-specific cumulative incidences is implemented in the \texttt{adjustedcif()} function. This function uses nearly equivalent syntax as the shown \texttt{adjustedsurv()} function, although fewer adjustment methods are implemented. More information is given in the package documentation.
\par
One limitation of this package is, that it currently only supports adjustment for baseline confounders. The presence of time-varying confounding introduces a lot of complicated issues for the estimation of causal effects, such as feedback between the treatment, confounders and mediators of the treatment and outcome relationship \parencite{Clare2019}. As such, it usually requires the usage of specialized g-methods, such as marginal structural models \parencite{Robins2000}, structural nested mean models \parencite{Vansteelandt2014} or forms of time-dependent matching \parencite{Thomas2020} Due to the structure of these methods, they can usually not be applied "off-the-shelves", which makes it difficult to implement them in a re-usable fashion. Some packages, such as the \texttt{ipw} R package \parencite{vanderWal2011}, do implement such functionality and may thus be more useful in such scenarios. Another limitation is that our package currently does not allow automatic estimation of causal interactions or effect modification. Although it is possible to use the package for such analyses, it is not a built-in feature, which would make it easier for users to perform such analyses.
\par
Previous versions of the package included various forms of targeted maximum likelihood based estimators \parencite{Stitelman2010, Cai2020, Rytgaard2023a}. These methods are generally doubly-robust, as AIPTW, but also naturally respect the probability bounds and always produce non-increasing estimates. Unfortunately, these methods had to be removed from the package, because the packages that we based our implementation on (\texttt{survtmle} \parencite{Benkeser2019} and \texttt{concrete} \parencite{Chen2023}) are no longer available on CRAN. If these or similar packages become available again in the future, we plan to include them in this package as well. Additionally, a few other methods based on covariate balancing have been proposed recently \parencite{Xue2023, GarciaMeixide2023, Pham2023}. We aim to include implementations of these methods in a future version of this package. 
\par
Furthermore, the \texttt{adjustedCurves} package only supports categorical treatment variables. If the variable of interest is continuous, the user has to artificially categorize it before applying any of the methods included in the package. This is generally inadvisable \parencite{Naggara2011}. In related work, we propose a possible solution to this problem \parencite{Denz2023}, which we implemented in the \texttt{contsurvplot} \textbf{R} package \parencite{Denz2023a}.

\section*{Acknowledgements}

We would like to thank Xiaofei Wang and Fangfang Bai for allowing us to incorporate their \textbf{R}-Code used for empirical likelihood estimation into our package directly. We also want to thank Jixian Wang for sending us their \textbf{R}-Code used for the calculation of the augmented inverse probability of treatment weighted estimator based on Pseudo-Values, which helped immensely to create the version implemented in our package. Additionally we want to thank Mirko Signorelli for his helpful comments and suggestions. Finally, we would like to thank both anonymous reviewers for their invaluable suggestions and corrections, which highly improved this manuscript.

\FloatBarrier
\newpage

\printbibliography

\end{document}